\documentclass[sigconf,nonacm=true]{acmart}

\renewcommand\footnotetextcopyrightpermission[1]{}
\setcopyright{none}

\newcommand{\datarelease}{at
  \url{https://huggingface.co/datasets/sajalregmi4/agent-town-economy}}

\usepackage{enumitem}

\graphicspath{{figures/}{../figures/}}

\ifdefined\compresslayout
  \newcommand{\widetab}[3]{%
    \begin{table}[t]%
      \centering
      \caption{#1}%
      \label{#2}%
      \small
      \resizebox{\linewidth}{!}{#3}%
    \end{table}}
\else
  \newcommand{\widetab}[3]{%
    \begin{table*}[t]%
      \centering
      \caption{#1}%
      \label{#2}%
      \small
      #3%
    \end{table*}}
\fi

\newcommand{\papertitle}{But How Would AI Agents Run a Town's Economy?}

\newcommand{\paperauthors}{%
  \author{Sajal Regmi}%
  \authornote{Primary contributor.}%
  \affiliation{\institution{Karela Technologies Inc.}\country{USA}}%
  \email{sajal@karelatechnologies.com}%
  \author{Siddhartha Pudasaini}%
  \affiliation{\institution{Karela Technologies Inc.}\country{USA}}%
  \author{Chetan Phakami Pun}%
  \affiliation{\institution{Karela Technologies Inc.}\country{USA}}%
}

\begin{document}

\title{\papertitle}
\paperauthors

\begin{abstract}
We placed 100 memory-equipped large language model (LLM) agents in
charge of a closed,
money-conserving spatial economy on real Pokhara Lakeside geography (earning wages,
running businesses, setting prices) and ran this multi-agent simulation for up to 26
simulated weeks, well past the 1--2 weeks typical of agent-society studies. Across 91 validated runs (2.44M agent
decisions, 21.5B tokens), the money stops moving, in a specific and
measurable way. A 12$\times$ tourist demand shock raises business
revenue 4.62$\times$ ($p<0.001$), which we decompose \emph{exactly}
into a 1.50$\times$ extensive margin (more businesses trading) and a
3.07$\times$ intensive margin (more revenue each). Monetary transmission stops there. Wages move 1.03$\times$ ($p=0.42$); 0.3\% of 3{,}981 menu
items are ever repriced ($p=0.47$). A randomized cash transfer (NPR
5{,}000 to 20 of 100 agents) shows the same pattern from the
opposite direction: 96.7\%
is still held 311 pulses later, marginal propensity to consume 3--4\%
by two independent measures, indistinguishable from zero. The wealth
distribution is consequently near-frozen at the horizon this
literature uses ($\rho=0.964$ over 2 simulated weeks), but not frozen. $\rho$ falls to 0.832 at 12 weeks and 0.752 at 26,
a horizon-dependence no short study can see. Matched ablations show
which knob actually matters. Swapping the backing LLM moves every
outcome we measure ($p=0.0039$); deleting agents' memory moves none
of them detectably. A purely social tool fails 94--97\% of the time
across two model families, compared with $\sim$96\% success on economic
tools, with no measurable shift away from it. Every headline number
is verified twice, by a live validator and by an offline recomputation
that reconciles each agent's wealth against its own signed
transaction history, and we release the full run corpus for
reanalysis.
\end{abstract}

\maketitle

\begin{center}
\emph{Extended version: not constrained to a conference page limit, with
Artifact Availability as its own section.}
\end{center}

  \section{Introduction}
\label{sec:intro}

A growing body of work embeds LLM agents in towns, markets, and
organizations and asks what happens
\cite{park2023generative,vezhnevets2023generative,horton2023large}.
Most run for days or a couple of simulated weeks, report that agents
behave
in recognizably human ways, and stop there. The question that
matters for deploying agentic AI in economic roles is whether
resources that enter such a system \emph{move}: whether demand
reaches wages, whether windfalls get spent, whether a population's
wealth is genuinely evolving or sitting still on the only timescale
anyone has checked.

We built a closed, money-conserving spatial economy of 100
memory-equipped LLM agents on real Pokhara Lakeside geography, working
shifts, running or patronizing 762 registered businesses, and
transacting under a live price system. Every rupee is accounted for:
total money equals the sum of agent balances at each of 41{,}328
recorded pulses, exactly, across all 91 validated runs
(\S\ref{sec:validation}). This lets us follow a shock or a transfer
rupee by rupee, and run long enough (up to 26 simulated weeks,
against the 1--2 weeks typical of comparable work) to ask whether the
wealth distribution we observe is stable or just unobserved for long
enough.

\textbf{Money enters and stops at the firm.} A 12$\times$ tourist
shock is delivered cleanly ($\delta=1.00$, $p_{\mathrm{adj}}<0.001$)
and revenue rises 4.62$\times$, which decomposes \emph{exactly} into a
1.50$\times$ extensive margin (more businesses trading) and a
3.07$\times$ intensive margin (more revenue where they do). Neither
reaches workers: wages move 1.03$\times$ ($p=0.42$); 0.3\% of
3{,}981 menu items are ever repriced ($p=0.47$). The chain a
textbook tourism economy predicts breaks at the second link.

\textbf{Money that reaches households does not circulate.} A
within-run randomized transfer (NPR 5{,}000 to 20 of 100 agents)
confirms this: 96.7\% is still held 311 pulses later, and
two independent MPC estimates agree at 3.3\% and 4.0\%, neither
distinguishable from zero.

\textbf{The distribution is frozen, but only if you stop early.} At the
conventional 2-week horizon, wealth persistence is $\rho=0.964$ and
only 21\% of agents ever change quintile. At 12 weeks (4 runs) $\rho$
falls to 0.832. In a single 26-week run, three times longer
than any comparable economic simulation we know of, it drops to 0.752, with the Gini
relaxing from 0.674 to 0.643 by week 16. Horizon changes the answer
from ``frozen'' to ``slowly equilibrating.''

Two further results complete the picture. Swapping the backing LLM
moves every outcome at the $p=0.0039$ floor available with 9 matched
seeds; removing agents' memory moves none of them detectably. And
\texttt{invite\_to\_talk} fails 94--97\% of the time across two model
families, accounting for 79.6\% of all tool failures.

\paragraph{Contributions.} (1) A validated, released dataset of 91
runs (2.44M decisions, 21.5B tokens), audited by two independent
pipelines (\S\ref{sec:validation}); (2) an identified mechanism for monetary
transmission failure, decomposed into extensive and intensive margins
(\S\ref{sec:transmission}) and corroborated by a randomized transfer
experiment (\S\ref{sec:grant}); (3) direct evidence that study
horizon changes the qualitative conclusion about wealth mobility
(\S\ref{sec:horizon}); and (4) a demonstration that model choice, not
memory, drives behavior, alongside a persistent, model-agnostic
failure of social coordination tooling (\S\ref{sec:ablations}). We
release the run corpus; the platform itself is not
(\S\ref{sec:artifact}).%
  \section{Related Work}
\label{sec:related}

\paragraph{Generative agent societies.} \citet{park2023generative}
found that LLM agents equipped with memory, reflection, and planning
exhibit credible individual and emergent social behavior across
roughly two simulated days, extending earlier persona-populated
interfaces \cite{park2022social}. Concordia
\cite{vezhnevets2023generative} broadens this into a reusable
grounding framework and has been deployed at social-network scale
\cite{gao2023s3}. Agentopia \cite{wang2026agentopia} pushes life
simulation to 10 simulated years with 100 agents, but its coarse time
units and well-being objectives leave economic aggregates unmeasured. This line of work demonstrates qualitatively
plausible individual behavior, but does not instrument a closed
resource system finely enough to test whether
aggregate outcomes (prices, wages, wealth) react to shocks the way a
real economy does, and horizons remain short. Our platform addresses
this directly: each transaction is ledgered, money is conserved and
can be independently audited (\S\ref{sec:validation}), and runs last
26 simulated weeks.

\paragraph{LLMs as economic agents.} A distinct strand examines
whether LLMs replicate human-subject behavior in surveys, auctions,
and bargaining \cite{horton2023large,argyle2023out}, or whether they
can trade autonomously \cite{li2023tradinggpt}. These studies usually
probe a single decision or one narrow mechanism, not 100 agents
jointly sustaining an economy over weeks. EconAgent
\cite{li2024econagent} and EconSimulacra
\cite{hashimoto2026econsimulacra} come closest, driving macroeconomic
and consumer-economy simulations with LLM agents; neither conserves
money at the transaction level, which is what licenses the exact
margin decomposition in \S\ref{sec:transmission}. Our results speak to this
literature's core benchmark, the marginal propensity to consume out
of a windfall, typically 0.2--0.5 for human households
\cite{johnson2006household,jappelli2014fiscal}. Our measured 3--4\%
is a sharp outlier against it.

\paragraph{Multi-agent benchmarks.} AgentBench
\cite{liu2024agentbench} and WebArena \cite{zhou2024webarena} score
agents on a per-episode success criterion. Our setting has no such
criterion; instead we follow the aggregate trajectory of a closed
economy across thousands of decision rounds, which is why we report
macroeconomic and distributional metrics (Gini
\cite{gini1912variabilita}, wealth-rank persistence, wage
pass-through) in place of task-level scores.

\paragraph{Positioning.} This work is an empirical
multi-agent-systems contribution, not a capability demonstration. We
aim to identify a mechanism behind a specific failure, not to show
that LLM agents can complete some task. The platform serves as the
apparatus; the dataset and the mechanism are what we contribute.
\citet{larooij2025critical} identify validation as the weak point of
generative agent-based modeling; the conserved-money ledger and two
independent validators (\S\ref{sec:validation}) are our answer to it.%
  \section{The Environment}
\label{sec:environment}

\begin{figure*}[t]
  \centering
  \includegraphics[width=\linewidth]{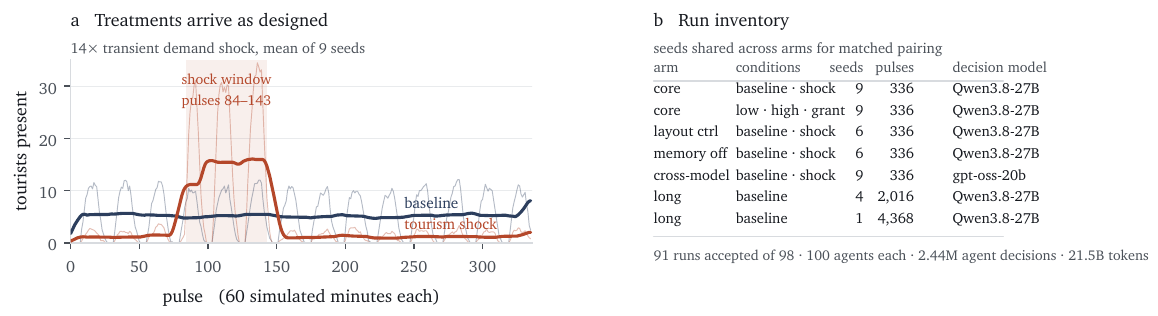}
  \caption{(a) The tourism-shock treatment: a transient
  14$\times$ rise in tourists present, mean of 9 seeds. (b) The 91
  accepted runs analyzed here, across seven arms that share a common
  seed pool for matched pairing.}
  \Description{Two panels. Panel a plots tourists present against pulse index for
  the baseline and tourism-shock conditions, averaged over 9 seeds;
  the shock condition rises about fourteen-fold during a shaded window
  spanning pulses 84 to 143 and returns to baseline afterwards, while
  the baseline condition stays flat. Panel b is a table of the 91
  accepted runs, listing for each of seven arms the conditions
  covered, the number of seeds, the pulses per run, and the decision
  model: core arms at 336 pulses on Qwen3.8-27B, layout-control and
  memory-off arms at 6 seeds, a cross-model arm on gpt-oss-20b, and
  two long arms at 2,016 and 4,368 pulses.}
  \label{fig:design}
\end{figure*}

\begin{figure}[t]
  \centering
  \includegraphics[width=0.78\linewidth]{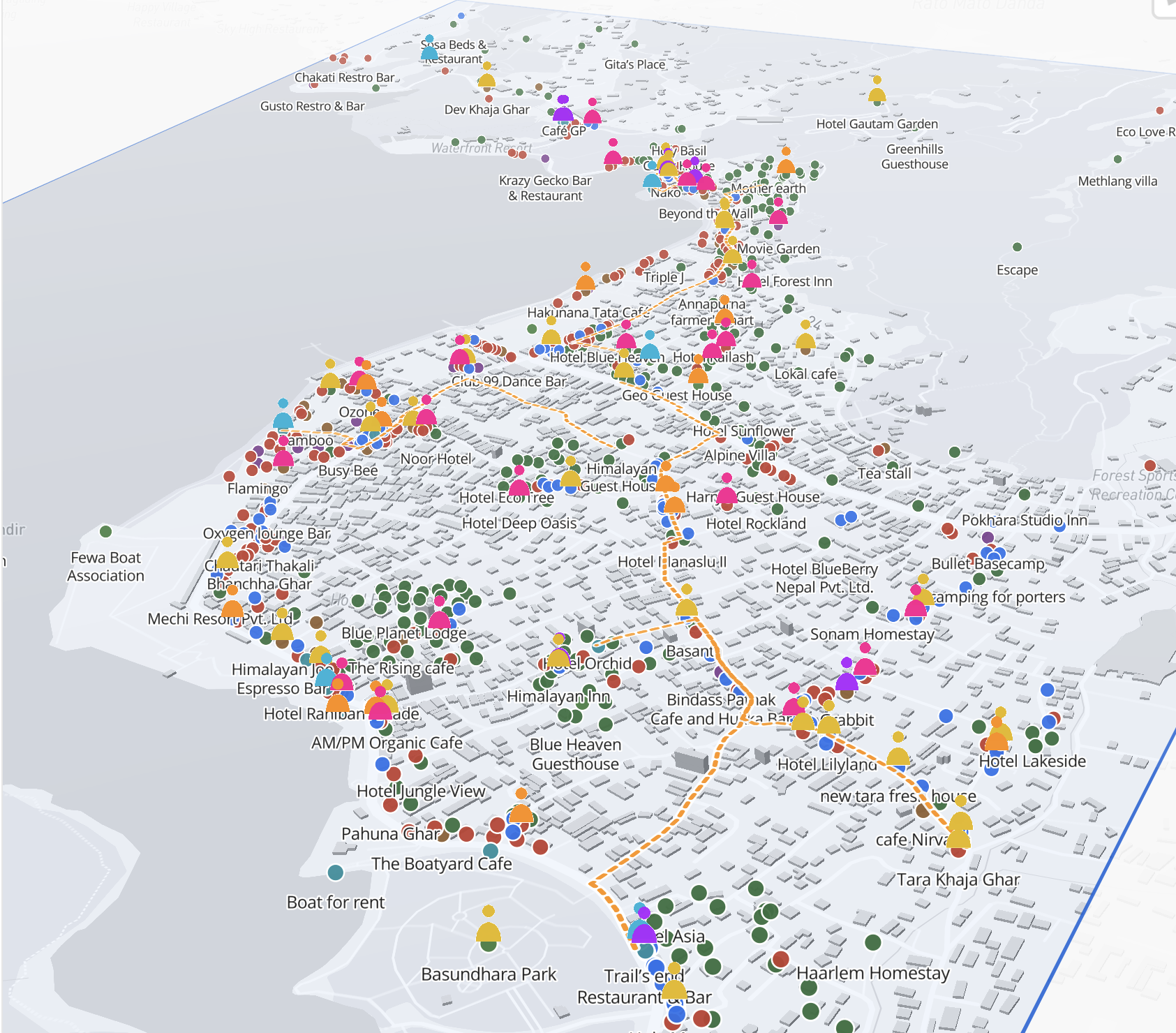}
  \caption{A run rendered over real Lakeside, Pokhara geography;
  markers are agents at their current place among 762 registered
  businesses. Map data \copyright{} OpenStreetMap, via Mapbox.}
  \Description{A screenshot of the simulation rendered over a street map of
  Lakeside, Pokhara, Nepal. Coloured circular markers scattered across
  the map show the current location of each of the 100 agents among
  the 762 registered businesses, which are labelled with real business
  names. The lake occupies the lower left of the frame. The map
  imagery is third-party material from OpenStreetMap via Mapbox.}
  \label{fig:setting-c}
\end{figure}

\textbf{Setting.} The world is modeled on Lakeside, Pokhara, Nepal,
with real building footprints and business locations
(Figure~\ref{fig:setting-c}). The registry holds 762 places across
seven categories; 735 carry priced menus totaling 3{,}981 items. One hundred LLM agents populate the town, each with a
cash wallet, a bank balance, and an 80\,m visibility radius gating
what it can perceive.

\textbf{Two clocks.} The \emph{world clock} advances in fixed
60-simulated-minute pulses, at which the platform resolves scheduled
dynamics and snapshots system state. The \emph{agent clock} is
event-driven: each agent wakes, reasons, and acts asynchronously
within a pulse rather than all 100 acting in lockstep. A 336-pulse
(2-week) run is a continuous, staggered stream of reasoning episodes
rather than 336 rounds of simultaneous decisions. The corpus totals
2.44M agent decisions and 21.5B tokens (\S\ref{sec:validation}).

\textbf{Agents and tools.} Every agent shares one specification: an
LLM decision policy (\S\ref{sec:design}), persistent memory (ablated
in one arm; \S\ref{sec:ablations}), and a fixed 19-action toolset:
navigation, the local economy (\texttt{buy\_food}, \texttt{set\_price},
\texttt{start\_shift}, \texttt{pay\_agent}), memory
(\texttt{write\_memory}, \texttt{search\_memory}, \texttt{reflect}),
and social coordination (\texttt{invite\_to\_talk},
\texttt{accept\_invite}, \texttt{leave\_conversation}). A subset own a
business and may set its prices and pay its wages. Every tool call,
transaction, price observation, and conversation turn is logged with
pulse-level timestamps.

\textbf{Money conservation.} Total money (every agent's wallet
plus bank balance) is conserved exactly except through two
ledgered external accounts, a tourist inflow pool and a government
sink. Money that leaves agent hands accrues to business cash-on-hand
and bank balances.%
  \section{Experimental Design}
\label{sec:design}

\textbf{Conditions and arms.} Every headline run uses the same baseline
agent specification (\S\ref{sec:environment}); they differ only in
world state and shocks. Five treatment conditions alter the tourist
arrival rate and two scheduled interventions: \textsc{baseline}
(2.0/sim-hour), \textsc{tourism-low} (0.5), \textsc{tourism-high}
(6.0), \textsc{tourism-shock} (arrival steps
low$\to$high$\to$low across pulses 84--143, Figure~\ref{fig:design}a),
and \textsc{wealth-grant} (baseline arrivals plus a transfer to 20 of
100 agents at pulse 24). Three more arms keep the condition fixed and
change one factor each: a duplicate world layout (\textsc{layout-b}),
agent memory removed (\textsc{no-memory}), and the decision LLM
replaced (\textsc{cross-model}). Two arms lengthen baseline to 12 and
26 simulated weeks. Figure~\ref{fig:design}b lists the full inventory.
Seeds are drawn from a single pre-registered pool shared across all
arms, which makes the matched-seed pairing used throughout
\S\ref{sec:ablations} possible.

\textbf{Deviations from pre-registration.} Conditions, seed lists, and
directional hypotheses were committed before data collection. The
planned headline sweep was 5 conditions $\times$ 30 seeds, with a
budget rule set in advance: fall back to 15 seeds/condition if per-run
cost exceeded projections. That call was made before any comparative
result was seen. The realized sweep fell below even that fallback: 9
seeds/condition, a 336-pulse horizon instead of 168, self-hosted
Qwen3.8-27B-FP8 instead of the planned MiMo, and ablations on
memory/layout/cross-model instead of agent count/visibility radius.
The cause was a mid-project shift to self-hosted inference, which
traded seed count for a longer horizon. We are explicit about this:
9 seeds is too few to rely on between-seed asymptotics, so every
headline comparison below is a matched-seed or within-run contrast.

\textbf{Exclusions.} Of 98 dispatched runs, 91 are used here. Six were
dropped from \textsc{cross-model} because its third model,
Mistral-Small-3.2-24B, emits duplicate tool-call identifiers within a
single reasoning episode, which the inference server rejects on the
follow-up turn; the circuit breaker halted each such run within 4
pulses. We keep this in \S\ref{sec:ablations} as a negative result on
multi-hop tool-calling robustness. One further run
(\textsc{tourism-low}, seed 10091) drew 56 tourist arrivals against a
pre-registered $3\sigma$ floor of 56.5, even though it completed all
336 pulses and passed every other check. The pre-committed rule, not
the analyst, decides that exclusion.%
  \section{Validation}
\label{sec:validation}

Every number in this paper comes from a single committed pipeline
that reads the archived run artifacts directly; nothing is
hand-computed. Two independent validators gate every run: the
platform's live TypeScript validator checks run-level integrity at
completion (invariants, provider health, pulse-series completeness),
and an offline Python validator recomputes every headline number a
second time, in a second language, from the raw CSV exports. Its most
demanding check is a per-agent ledger reconciliation: the change in
wallet-plus-bank balance must equal the signed sum of that agent's
own transaction history, exactly, which catches export truncation and
accounting bugs a purely aggregate check would miss.

\begin{table}[t]
  \centering
  \caption{Independent Python revalidation from raw CSV exports, all
  91 accepted runs. Full 19-check breakdown released with the data.}
  \label{tab:validation}
  \small
  \begin{tabular}{p{0.72\linewidth}r}
\toprule
Check & Runs passing \\
\midrule
17 checks (\texttt{ledger\_vs\_panel}, \texttt{money\_series}, all \texttt{flow\_*} totals, \ldots) & 91/91 each \\
\addlinespace
\texttt{terminal\_gini\_vs\_metrics} (warn) & 28/91 \\
\texttt{tokens\_vs\_ledger} (fail) & 73/91 \\
\bottomrule
\end{tabular}
\end{table}

In Table~\ref{tab:validation}, 17 of 19 checks, including
the ledger reconciliation and every transaction-flow total, pass
on all 91 runs with zero exceptions. \texttt{terminal\_gini\_vs\_metrics}
(28/91) flags a benign staleness issue: the live \texttt{metrics.json}
snapshot reads current world state at export time and can drift
$\sim$0.001 in Gini if another run later ran on the same world; the
pulse-level CSV panel used throughout this paper is unaffected.
\texttt{tokens\_vs\_ledger} (73/91) fails on every gpt-oss run and
no Qwen run: gpt-oss forfeits a small share of malformed generations
(\S\ref{sec:ablations}) counted in the pulse-level token counter but
excluded from the per-call ledger.

\textbf{Money conservation, directly.} The strictest check does not
appear in the table. \texttt{totalMoneyInSystem} equals the
independently computed sum of every agent's wallet and bank balance at
\emph{every one of the 41{,}328 recorded pulses across all 91 runs},
zero deviation. Money enters only through a ledgered tourist-inflow
account and leaves only through a ledgered government sink; every
rupee in between sits in an agent balance or a business's
cash-on-hand/bank balance. The margin decomposition in
\S\ref{sec:transmission} and the transfer-tracing in
\S\ref{sec:grant} both depend on this identity.

\textbf{Serving differences by model.} Table~\ref{tab:inference}
reports raw LLM call outcomes. Getting gpt-oss to a usable failure
rate required three changes, disclosed because they affect
\S\ref{sec:ablations}'s cross-model comparison: raising the output
budget 512$\to$1{,}024 tokens (gpt-oss reasons before a tool call;
cut malformed output 24.0\%$\to$6.0\%); disabling FP8 KV-cache
quantization (gpt-oss is natively quantized and produced garbage
tokens stacked with a second layer, 6.0\%$\to$3.3\%); and a tool-name
normalization step after the serving stack's parser leaked internal
channel markers into tool names, which had corrupted 16--35\% of
actions before the fix. A residual 0.86\% of gpt-oss calls still
return a malformed generation and are forfeited, against 0.00\% for
Qwen. Any tool-reliability probe on this platform must replay recorded
production requests; a reduced probe we ran early at 3 tools and
temperature 0 reported 100\% success on this same failure mode and
would have concealed the \S\ref{sec:ablations} result.

\widetab{LLM call outcomes by decision model, all accepted 336-pulse runs. \texttt{tool\_calls}/\texttt{length} are OpenAI-style finish reasons; malformed generations are forfeited, not retried.}%
  {tab:inference}{\begin{tabular}{lrrrrr}
\toprule
Decision model & Runs & LLM calls & \texttt{tool\_calls} & \texttt{length} & malformed \\
\midrule
\texttt{Qwen/Qwen3.8-27B-FP8} & 73 & 4,125,699 & 70.3\% & 15.2\% & 0.00\% \\
\texttt{openai/gpt-oss-20b} & 18 & 957,828 & 65.8\% & 26.7\% & 0.86\% \\
\bottomrule
\end{tabular}}%
  \section{Transmission Failure}
\label{sec:transmission}

\begin{figure*}[t]
  \centering
  \includegraphics[width=\linewidth]{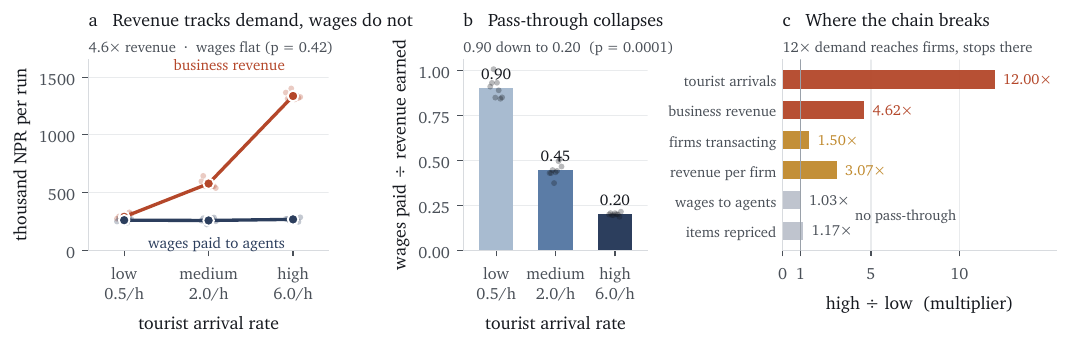}
  \caption{(a) Revenue tracks tourist arrival rate; wages do not. (b)
  Wage share of revenue collapses 0.90 (low) $\to$ 0.20 (high). (c)
  The multiplier cascade: 4.62$\times$ revenue decomposes exactly
  into a 1.50$\times$ extensive and 3.07$\times$ intensive margin,
  then stops, with wages and prices within noise of 1$\times$.}
  \Description{Three panels. Panel a plots thousands of NPR per run against tourist
  arrival rate at low 0.5, medium 2.0, and high 6.0 per hour: business
  revenue rises steeply from about 290 to about 1,340 thousand, while
  wages paid to agents stay flat near 260 thousand across the whole
  range. Panel b shows wages as a share of revenue falling
  monotonically across the same three arrival rates, from 0.90 to
  0.45 to 0.20. Panel c is a horizontal bar chart of high-divided-by-low
  multipliers: tourist arrivals 12.00 times by design, business
  revenue 4.62 times, firms transacting 1.50 times, revenue per firm
  3.07 times, and then wages to agents 1.03 times and menu items
  repriced 1.17 times, the last two indistinguishable from a
  multiplier of one.}
  \label{fig:transmission}
\end{figure*}

We use the tourism-intensity sweep, spanning \textsc{tourism-low}
(0.5/sim-hour), \textsc{baseline} (2.0), and \textsc{tourism-high}
(6.0), as a controlled comparative-statics experiment. The manipulation
check is unambiguous (Table~\ref{tab:tourism-sweep}): tourist spending
separates completely ($\delta=1.00$, $p_{\mathrm{adj}}<0.001$), so any
downstream null cannot be attributed to a weak treatment.

\widetab{Tourism intensity sweep, terminal-pulse outcomes (mean over 8--9 seeds). $\delta$: Cliff's delta, high vs.\ low, Holm-Bonferroni-adjusted across the six-outcome family.}%
  {tab:tourism-sweep}{\begin{tabular}{lrrrrr}
\toprule
& \multicolumn{3}{c}{Tourist arrival rate (per sim-hour)} & \multicolumn{2}{c}{high vs low} \\
\cmidrule(lr){2-4}\cmidrule(lr){5-6}
Outcome & 0.5 & 2.0 & 6.0 & $\delta$ & $p_{\mathrm{adj}}$ \\
\midrule
Gini (terminal) & 0.6712 & 0.6716 & 0.6706 & -0.28 & 0.994 \\
Median wealth (NPR) & 23,954 & 24,307 & 24,659 & +0.50 & 0.278 \\
Purchase attempts & 1,262 & 1,317 & 1,289 & +0.14 & 1.000 \\
Purchase failure rate & 0.1300 & 0.1291 & 0.1279 & +0.03 & 1.000 \\
Tourist spending (NPR) & 87,639 & 370,576 & 1,133,633 & +1.00 & $<$0.001 \\
Persistence $\rho$ & 0.9702 & 0.9643 & 0.9683 & -0.33 & 0.416 \\
\bottomrule
\end{tabular}}

\textbf{Revenue responds; wages and prices do not.} Business revenue
rises 4.62$\times$ across the sweep ($p=0.0001$; Figure~\ref{fig:transmission}a),
and place-level wealth rises with it, 9.8M$\to$14.9M NPR. Two
outcomes that a standard demand-pass-through model predicts would
move in step do not move at all. Wages are flat across the full
12$\times$ range: 260{,}389 vs.\ 267{,}459 NPR ($p=0.42$). The wage
share of revenue falls monotonically, 0.90 (low) $\to$ 0.45
(baseline) $\to$ 0.20 (high; Figure~\ref{fig:transmission}b). As the
tourist economy grows, a shrinking share of the money it generates
reaches the agents staffing it. Prices are more static still: of
3{,}981 menu items, only 0.27--0.32\% are ever repriced, unresponsive
to a 4.6$\times$ revenue swing ($p=0.47$). Terminal median wealth is not distinguishable from noise
($p_{\mathrm{adj}}=0.28$).

\textbf{The chain breaks at the firm.} Money conservation
(\S\ref{sec:validation}) lets us decompose the 4.62$\times$ response
exactly, not just observe it. The \emph{extensive margin} is how many
of 762 registered businesses record any transaction; the
\emph{intensive margin} is revenue per business that does.
Table~\ref{tab:margins} and Figure~\ref{fig:transmission}c report
both: 1.50$\times$ more businesses trade at high demand (458 vs.\ 304
of 762, $p<0.001$), and each earns 3.07$\times$ more ($p<0.001$).
These multiply to $1.504\times3.067=4.61$ against an observed 4.62,
matching up to rounding. Demand growth is absorbed by drawing
more of the fixed registry into activity and by deepening takings at
active businesses. Neither channel has any mechanical connection to a
wage or a menu price, and empirically, neither moves one.

\begin{table}[t]
  \centering
  \caption{The multiplier cascade, low vs.\ high tourist arrival.}
  \label{tab:margins}
  \small
  \begin{tabular}{lrrrl}
\toprule
Quantity & low (0.5/h) & high (6.0/h) & Ratio & $p$ \\
\midrule
Tourist arrival rate & 0.5 & 6.0 & 12.00$\times$ & by design \\
Business revenue (NPR) & 289,370 & 1,337,007 & 4.62$\times$ & $<$0.001 \\
\addlinespace
\quad firms transacting (of 762) & 304 & 458 & 1.50$\times$ & $<$0.001 \\
\quad revenue per firm (NPR) & 953 & 2,922 & 3.07$\times$ & $<$0.001 \\
\addlinespace
Wages paid to agents (NPR) & 260,389 & 267,459 & 1.03$\times$ & 0.423 \\
Menu items repriced & 0.27\% & 0.32\% & -- & 0.468 \\
\bottomrule
\end{tabular}
\end{table}

\textbf{Interpretation.} We do not claim agents are irrational.
\texttt{set\_price} is called only $\sim$18 times per run out of
$\sim$28{,}000 total calls, so this is partly about agents rarely
exercising a lever that exists, and partly about how sticky the
initial menu is by construction (\S\ref{sec:limitations}). In a
closed, audited economy where agents \emph{can} adjust wages and
prices, a cleanly-identified demand shock does not propagate past the
firm under the policies agents actually adopt: gains concentrate at
owners instead of passing to workers or customers.%
  \section{Windfalls Are Not Circulated}
\label{sec:grant}

\begin{figure*}[t]
  \centering
  \includegraphics[width=\linewidth]{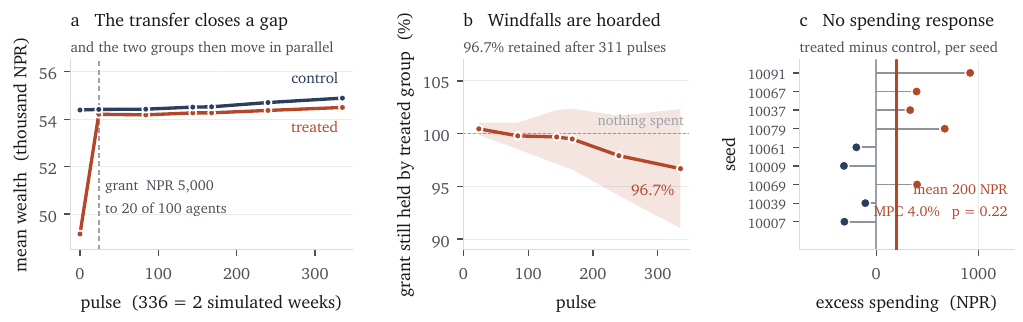}
  \caption{(a) Mean wealth, treated (+NPR 5{,}000 at pulse 24) vs.\
  control: the transfer closes the gap instantly, then both move in
  parallel. (b) Share of the transfer still held over time (s.e.\
  band, 9 seeds). (c) Excess purchase spending per seed; the mean is
  not distinguishable from zero.}
  \Description{Three panels. Panel a plots mean wealth in thousands of NPR against
  pulse for treated and control agents: the treated line starts about
  5,000 NPR below control, jumps to meet it at pulse 24 when the grant
  is paid, and the two lines then run in parallel for the remaining
  311 pulses. Panel b plots the percentage of the transfer still held
  by the treated group over time with a standard-error band across 9
  seeds, declining only slightly to 96.7 percent by the end of the
  run. Panel c is a dot plot of excess purchase spending in NPR for
  each of nine seeds, scattered on both sides of zero, with a mean of
  about 200 NPR marked by a vertical line, corresponding to a marginal
  propensity to consume of 4.0 percent at p equals 0.22.}
  \label{fig:grant}
\end{figure*}

\S\ref{sec:transmission} shows money does not reach households
through the firm. We now ask the complementary question, under the
strongest identification available here: if money \emph{does} reach a
household directly, does it move? We answer with a cash transfer
randomized \emph{within} a single run. Treated and control share
the same world, shocks, and counterparties, so the comparison needs
no cross-seed variance assumption. In \textsc{wealth-grant}, 20 of
100 agents are chosen at random and receive NPR 5{,}000 at pulse 24
from the ledgered government account; the remaining 80 form the
control. We track the wealth gap across all 336 pulses and
independently measure post-transfer purchase spending, for two
estimates of the marginal propensity to consume (MPC).

\textbf{The transfer lands and stops moving.} The treated--control gap
jumps from $-$NPR 5{,}229 (pre-transfer imbalance) to $+$5{,}023 at
pulse 24, essentially the full grant ($t=176.8$, $p<10^{-14}$;
Figure~\ref{fig:grant}a), then decays only slowly: 99.8\% still held
at pulse 84, 96.7\% at pulse 335 ($t=17.1$, $p<10^{-6}$;
Figure~\ref{fig:grant}b), no sign of reversion. We cross-check against
actual spending: treated agents spend a mean of NPR 2{,}096
post-transfer against 1{,}896 for control, an excess of 200
(Figure~\ref{fig:grant}c) that is not distinguishable from zero
($p=0.22$ $t$-test, $p=0.16$ Wilcoxon). The two MPC estimates agree: 3.3\% from
wealth trajectory, 4.0\% from spending, both far below the 0.2--0.5
range for human households \cite{johnson2006household,jappelli2014fiscal}.

\textbf{Interpretation.} These agents are, in aggregate, savers
rather than spenders. A windfall that does
not circulate cannot generate the second-round demand that would let
it reach anyone else, part of why the distribution in
\S\ref{sec:horizon} moves so slowly.%
  \section{The Wealth Distribution Only Looks Frozen}
\label{sec:horizon}

\begin{figure*}[t]
  \centering
  \includegraphics[width=\linewidth]{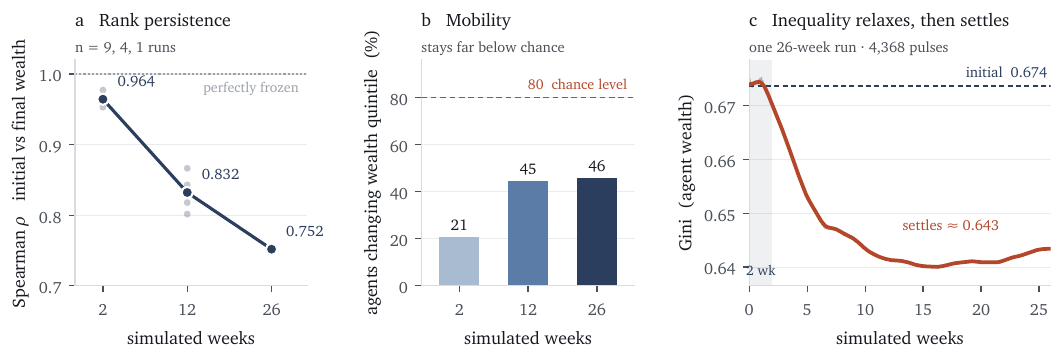}
  \caption{(a) Spearman correlation between initial and terminal
  wealth vs.\ run length. (b) Share of agents changing wealth
  quintile; dashed line marks the 80\% expected under independent
  reshuffle. (c) Gini across the single 26-week run: a fast initial
  rise, then slow relaxation to a plateau near week 16.}
  \Description{Three panels. Panel a plots the Spearman correlation between initial
  and terminal wealth against run length in simulated weeks, falling
  from 0.964 at 2 weeks with 9 runs, to 0.832 at 12 weeks with 4 runs,
  to 0.752 at 26 weeks with a single run; a dashed line at 1.0 marks a
  perfectly frozen distribution. Panel b is a bar chart of the
  percentage of agents changing wealth quintile at the same three
  horizons, rising from 21 to 45 to 46 percent, all far below a marked
  80 percent chance level. Panel c plots the Gini coefficient of agent
  wealth across the single 26-week run: it starts at 0.674, falls
  fairly sharply over the first ten weeks, and settles near 0.643 from
  about week 16 onward.}
  \label{fig:horizon}
\end{figure*}

\S\ref{sec:transmission}--\ref{sec:grant} identify a mechanism for
why money does not move. This asks what it implies for the wealth
distribution, and the answer depends on how long you watch.

\textbf{Persistence at the conventional horizon.} Over the 336-pulse
(2-week) baseline runs common in this literature, terminal wealth is
almost fully predictable from pulse-0 wealth: Spearman $\rho=0.964$
(Table~\ref{tab:horizon}), and only 21\% of agents ever change wealth
quintile, against 80\% expected under independent reshuffle. It holds across the tourism sweep ($\rho$ 0.964--0.970).
A study that runs two weeks and stops, which is most of the
generative-agent-society literature (\S\ref{sec:related}), would
correctly report this as static.

\textbf{It is not static.} We extend baseline to 12 weeks (4 runs)
and one run to 26, three times longer than any run in the headline
sweep. Persistence falls monotonically: $\rho=0.964$ at 2 weeks,
$0.832$ at 12, $0.752$ at 26 (Figure~\ref{fig:horizon}a), and quintile
mobility rises correspondingly, 21\%$\to$45\%$\to$46\%
(Figure~\ref{fig:horizon}b). Every 12-week run shows lower persistence
than every 2-week run; the ordering is not a mean-comparison
artifact. The single 26-week trajectory (Figure~\ref{fig:horizon}c)
makes the mechanism visible: Gini rises slightly in the first two
weeks, falls sharply to $\approx$0.641 by week 10, and plateaus near
0.643. The economy is
\emph{relaxing} on a timescale of roughly ten weeks, invisible to any
study run at the horizon this literature typically uses.

\widetab{Wealth-rank persistence by horizon, \textsc{baseline} condition. Runs at 12/26 weeks are few; see \S\ref{sec:limitations}.}%
  {tab:horizon}{\begin{tabular}{rrrrrrr}
\toprule
Sim.\ weeks & Pulses & Runs & $\rho$ & Quintile moved & Gini$_T$ & Top-10\% share \\
\midrule
2 & 336 & 9 & 0.964 & 21\% & 0.6716 & 0.5767 \\
12 & 2,016 & 4 & 0.832 & 45\% & 0.6543 & 0.5410 \\
26 & 4,368 & 1 & 0.752 & 46\% & 0.6434 & 0.5116 \\
\bottomrule
\end{tabular}}

\textbf{What this means for study design.} A wealth-distribution
conclusion drawn from an LLM-agent economy is only as trustworthy as
the
horizon it was measured over. The 12/26-week results rest on 5 runs, and the 26-week trajectory is a single path (\S\ref{sec:limitations}).
Still, the near-total separation between horizons is a stronger signal than
the raw run count suggests, and is, to our knowledge, the first
direct evidence in this literature that a stated conclusion about
wealth mobility depends on how long the simulation runs.%
  \section{What Actually Drives Behavior}
\label{sec:ablations}

\ifdefined\compresslayout
\begin{figure*}[t]
  \centering
  \includegraphics[width=\linewidth]{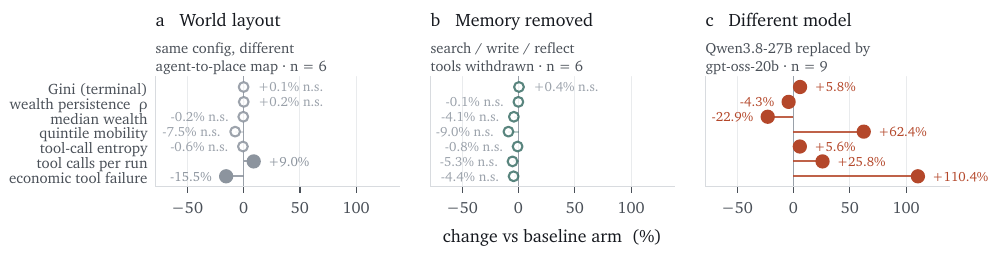}\\[4pt]
  \includegraphics[width=\linewidth]{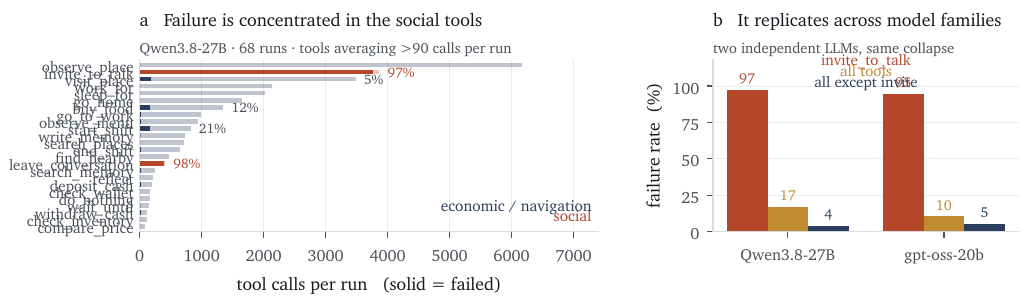}
  \caption{\emph{Top:} matched-seed contrasts vs.\ \textsc{baseline
  core-a} (filled = $p<0.05$). (a) Duplicate layout, a placebo. (b)
  Memory withdrawn. (c) LLM swapped for gpt-oss-20b, with an axis
  break; effects 3--10$\times$ larger. \emph{Bottom:} per-tool failure
  rate. (d) Qwen3.8-27B: social tools fail 94--100\%,
  economic/navigation mostly succeed. (e) Replicates almost unchanged
  in gpt-oss-20b.}
  \Description{Five panels in two stacked groups. The top group shows
  matched-seed contrasts against the baseline arm for eight outcomes,
  with filled markers indicating p below 0.05. Panel a, a duplicate
  world layout used as a placebo with 6 matched seeds, shows all
  outcomes within a few percent of zero. Panel b, memory tools
  withdrawn with 6 matched seeds, shows every outcome
  non-significant, from plus 0.4 to minus 9.0 percent. Panel c, the
  decision model swapped to gpt-oss-20b with 9 matched seeds, uses a
  broken axis and shows every outcome significant and three to ten
  times larger, including Gini plus 5.8 percent, median wealth minus
  22.9 percent, and economic tool failure plus 110.4 percent. The
  bottom group shows per-tool failure rates. Panel d, under
  Qwen3.8-27B, is a horizontal bar chart of tool calls per run with
  the failed portion shown solid: invite_to_talk fails 97 percent and
  leave_conversation 98 percent, while economic and navigation tools
  fail around 5 to 21 percent. Panel e shows the same pattern
  replicating almost unchanged under gpt-oss-20b.}
  \label{fig:ablations}
\end{figure*}
\else
\begin{figure*}[t]
  \centering
  \includegraphics[width=\linewidth]{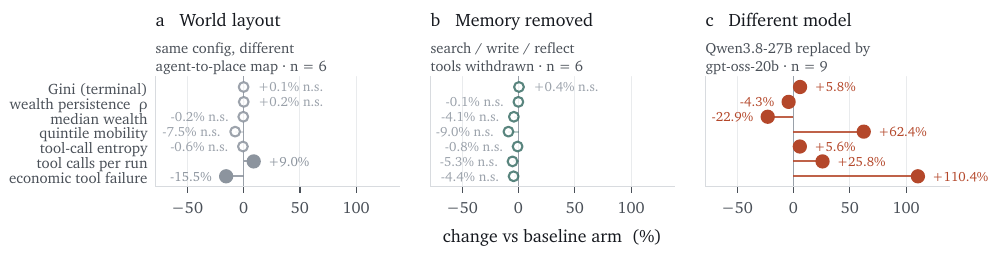}
  \caption{Matched-seed contrasts vs.\ \textsc{baseline core-a}
  (filled = $p<0.05$). (a) Duplicate layout, a placebo. (b) Memory
  withdrawn. (c) LLM swapped for gpt-oss-20b, with an axis
  break; effects 3--10$\times$ larger.}
  \Description{Three panels of matched-seed contrasts against the baseline arm,
  each plotting percentage change for eight outcomes, with filled
  markers indicating p below 0.05 and hollow markers indicating no
  significant difference. Panel a, a duplicate world layout used as a
  placebo with 6 matched seeds, shows all outcomes within a few
  percent of zero and non-significant except tool calls per run at
  plus 9.0 percent and economic tool failure at minus 15.5 percent.
  Panel b, memory tools withdrawn with 6 matched seeds, shows every
  outcome non-significant, ranging from plus 0.4 to minus 9.0 percent.
  Panel c, the decision model swapped to gpt-oss-20b with 9 matched
  seeds, uses a broken axis and shows every outcome significant and
  far larger, including Gini plus 5.8 percent, median wealth minus
  22.9 percent, quintile mobility plus 62.4 percent, and economic tool
  failure plus 110.4 percent.}
  \label{fig:ablations}
\end{figure*}

\begin{figure*}[t]
  \centering
  \includegraphics[width=\linewidth]{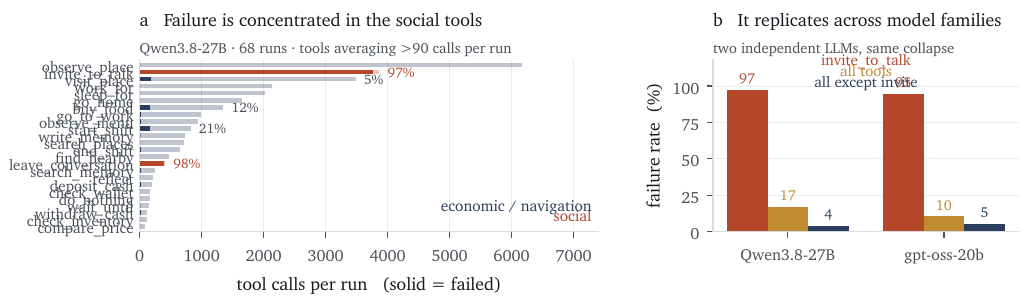}
  \caption{Per-tool failure rate. (d) Qwen3.8-27B: social tools fail
  94--100\%, economic/navigation mostly succeed. (e) Replicates
  almost unchanged in gpt-oss-20b.}
  \Description{Two panels on per-tool reliability. Panel a is a horizontal bar
  chart of tool calls per run under Qwen3.8-27B across 68 runs, with
  the failed portion of each bar shown solid: observe_place is the
  most called tool and mostly succeeds, invite_to_talk is second most
  called and fails 97 percent of the time, leave_conversation fails 98
  percent, while economic and navigation tools such as buy_food,
  go_to_work, and start_shift fail around 5 to 21 percent. Panel b is
  a grouped bar chart comparing failure rates between Qwen3.8-27B and
  gpt-oss-20b for invite_to_talk, all tools, and all tools except
  invite: the invite failure rate is 97 and 95 percent respectively,
  nearly identical across the two model families, while the
  all-except-invite rate stays at 4 to 5 percent for both.}
  \label{fig:tools}
\end{figure*}
\fi

The results so far describe \emph{what} the economy does. Here we ask
what, among the factors we control, actually determines it, using
three matched-seed contrasts against \textsc{baseline core-a}
(Figure~\ref{fig:ablations}, Table~\ref{tab:ablations}).

\widetab{Matched-seed contrasts vs.\ \textsc{baseline core-a}. $\Delta\%$: relative change; $p$ from Wilcoxon signed-rank ($n{=}6$ layout/memory, $n{=}9$ cross-model).}%
  {tab:ablations}{\begin{tabular}{l rr rr rr}
\toprule
& \multicolumn{2}{c}{Layout B} & \multicolumn{2}{c}{Memory off} & \multicolumn{2}{c}{gpt-oss-20b} \\
\cmidrule(lr){2-3}\cmidrule(lr){4-5}\cmidrule(lr){6-7}
Outcome & $\Delta\%$ & $p$ & $\Delta\%$ & $p$ & $\Delta\%$ & $p$ \\
\midrule
Gini (terminal) & +0.1 & 1.000 & +0.4 & 0.562 & +5.8 & 0.004 \\
Persistence $\rho$ & +0.2 & 0.844 & -0.1 & 0.688 & -4.3 & 0.004 \\
Median wealth (NPR) & -0.2 & 0.688 & -4.1 & 0.156 & -22.9 & 0.004 \\
Quintile mobility & -7.5 & 0.688 & -9.0 & 0.562 & +62.4 & 0.004 \\
Tool-call entropy & -0.6 & 0.844 & -0.8 & 0.438 & +5.6 & 0.004 \\
Tool calls per run & +9.0 & 0.031 & -5.3 & 0.062 & +25.8 & 0.004 \\
Economic tool failure & -15.5 & 0.031 & -4.4 & 0.438 & +110.4 & 0.004 \\
\texttt{invite\_to\_talk} failure & -0.3 & 0.688 & -2.2 & 0.031 & -2.8 & 0.004 \\
\bottomrule
\end{tabular}}

\textbf{A placebo: world layout.} \textsc{core-b} reuses two of six
worlds from a second, independently generated agent-to-place layout,
motivating a placebo arm (\textsc{layout-b}) holding every parameter
and seed fixed and varying only the layout. It does not: Gini
$+0.1\%$ ($p=1.00$), persistence $+0.2\%$ ($p=0.84$), all other
outcomes similarly null, which licenses \textsc{core-a} and
\textsc{core-b} as one population throughout.

\textbf{Memory: no detectable effect.} \textsc{no-memory} withdraws
\texttt{write\_memory}, \texttt{search\_memory}, \texttt{reflect}
entirely, verified by zero recorded calls against 1{,}369/run in
the control. All economic outcomes are statistically indistinguishable
from baseline: Gini $+0.4\%$ ($p=0.56$), persistence $-0.1\%$
($p=0.69$), median wealth $-4.1\%$ ($p=0.16$). Only 6 matched seeds;
a null at that size means we \emph{detected} no effect, not that none
exists. The one outcome that does move is mechanical:
\texttt{invite\_to\_talk} failure falls $2.2\%$ ($p=0.031$),
plausibly because agents without memory cannot recall which partners
recently declined.

\textbf{Model identity: everything moves.} \textsc{cross-model} holds
environment, tools, and population fixed and swaps Qwen3.8-27B-FP8
for gpt-oss-20b. Every outcome moves, at the $p=0.0039$ floor
available to a 9-seed matched Wilcoxon: Gini $+5.8\%$, median wealth
$-22.9\%$, quintile mobility $+62.4\%$, economic tool failure
$+110.4\%$, which is 3--10$\times$ larger than the memory-ablation
effect. Unlike memory or layout, each clears significance. Which LLM
decides is a first-order determinant of the economy; whether it has
memory, among what we measure, is not.

\textbf{The social layer fails, model-agnostically.}
\label{sec:social}
\texttt{invite\_to\_talk} fails 97.2\% under Qwen, 94.8\% under
gpt-oss, second most frequently called tool after
\texttt{observe\_place}. \texttt{accept\_invite} fails 100.0\% of
1{,}413 calls; \texttt{leave\_conversation} 97.7\%. These three
account for 79.6\% of all tool failures; excluding them,
failure across remaining tools is 4.0\%. The dominant logged reason is
\texttt{"you are already in too many conversations"}, a platform
concurrency cap that occurred 3{,}894 times in one sampled run. The cap
is our design choice, not evidence about LLM social cognition. Agents
show no measurable adaptation to it, receiving this exact,
human-readable error thousands of times with no detectable shift away
from that tool, against the 96\% success rate of economic tools under
the same agents, context, and memory.%
  \section{Discussion}
\label{sec:discussion}

Four independently identified results point in the same direction:
demand reaches firms but not households (\S\ref{sec:transmission});
direct transfers do not circulate (\S\ref{sec:grant}); the wealth
distribution is near-static at conventional horizons only because
those horizons are too short to see it relax (\S\ref{sec:horizon});
and the one factor governing all of it is which LLM decides, not
memory (\S\ref{sec:ablations}). Each rests on its own design, and together they describe one failure
mode, in which agents earn, hold, and do not spend money at
every scale we tested.

If the goal is an economy that behaves like a plausible model of a
human one, this transmission failure is the first thing to fix, ahead
of memory or governance. Without an incentive to move wages or prices
when a shift sells out, this class of system concentrates windfalls
at firms by construction, regardless of the underlying LLM's
capability. \S\ref{sec:horizon} pushes the point wider: any
distributional claim from an LLM-agent society should report its
horizon, and a short-horizon ``stable'' finding may be an artifact of
stopping early. \S\ref{sec:social} narrows to a specific case: an
aggregate success rate can hide a near-total, model-agnostic failure
of one action class behind a healthy overall number.

Both headline claims are falsifiable from data we release. The
transmission-failure claim would fall to a prompting change, such as
reminding owners that raising wages is a live option when a shift
sells out, if that restored pass-through with nothing else altered.
We did not test this. The horizon claim would fall to 12/26-week
replicates that do not reproduce the ordering in
Table~\ref{tab:horizon}.%
  \section{Limitations}
\label{sec:limitations}

\begin{itemize}[leftmargin=*,itemsep=1pt,topsep=2pt]
\item \textbf{Long-horizon sample size.} \S\ref{sec:horizon} rests on
  9 runs at 2 weeks, 4 at 12, and \emph{one} at 26, a single path
  rather than a distributional estimate. The ordering claim (every
  12-week run below every 2-week run) is what we lean on.
\item \textbf{Underpowered memory null.} Only 6 matched seeds
  (\S\ref{sec:ablations}); ``no detected effect'' is not equivalence.
\item \textbf{Price stickiness is partly by construction.}
  \texttt{set\_price} is called $\sim$18 times/run; the near-zero
  repricing rate reflects rare use of an available lever \emph{and}
  a sticky menu by design; we have not separated the two.
\item \textbf{One town, one seed pool, one stack, no governance.} 9
  seeds vary agent draws on one map only, both models on self-hosted
  vLLM (\S\ref{sec:validation}); Mistral-Small-3.2-24B could not
  complete a single pulse (\S\ref{sec:design}), a stack-specific
  result rather than a general model claim.
\end{itemize}

\ifdefined\compresslayout
\smallskip
\noindent\textbf{Artifact availability.}
\label{sec:artifact}
All 91 accepted and 7 quarantined runs (\S\ref{sec:design}), with the
pipeline behind every number, are on Hugging Face under CC~BY~4.0
(data) and MIT (code), \datarelease. The platform is not released.
\fi%
  \ifdefined\compresslayout\else\section{Artifact Availability}
\label{sec:artifact}

We release the full validated run corpus described throughout this
paper: all 91 accepted runs and the 7 quarantined runs kept as
evidence (\S\ref{sec:design}), spanning 2.44M agent decisions and
21.5B tokens across 41{,}328 recorded pulses. The release is
CSV-level (per-pulse world state, per-agent wealth time series, every
transaction, tool call, price observation, and conversation turn),
together with the analysis pipeline that turns those raw
exports into every number, table, and figure in this paper
(\S\ref{sec:validation}), so no result here needs to be taken on
faith. The data is released under CC~BY~4.0 and the analysis code
under the MIT license, on Hugging Face \datarelease. The simulation
platform itself is not released.
\fi

\bibliographystyle{ACM-Reference-Format}
\bibliography{refs}

\end{document}